\documentclass[aps,prl,letterpaper,twocolumn,preprintnumbers,floatfix,superscriptaddress]{revtex4}

\usepackage{graphicx}
\usepackage{xcolor}
\usepackage{dcolumn}
\usepackage{bm}
\usepackage{epsfig,cancel}
\usepackage{amsmath}
\usepackage{amssymb}
\usepackage{graphicx,bm}

\def\fun#1#2{\lower3.6pt\vbox{\baselineskip0pt\lineskip.9pt
  \ialign{$\mathsurround=0pt#1\hfil##\hfil$\crcr#2\crcr\sim\crcr}}}

\def\lsim{\mathrel{\rlap{\raise 2.5pt \hbox{$<$}}\lower 2.5pt\hbox{$\sim$}}}
\def\gsim{\mathrel{\rlap{\raise 2.5pt \hbox{$>$}}\lower 2.5pt\hbox{$\sim$}}}

\input epsf

\usepackage{mathrsfs}
\usepackage{hyperref}
\usepackage[utf8]{inputenc}

\DeclareSymbolFont{bbold}{U}{bbold}{m}{n}
\DeclareSymbolFontAlphabet{\mathbbold}{bbold}

\begin{document}
\title{Gravitational wave and electroweak baryogenesis with two Higgs doublet models}

\author{Ruiyu Zhou}
\affiliation{School of Science, Chongqing University of Posts and Telecommunications, Chongqing 400065, P. R. China}
\author{Ligong Bian\footnote{Corresponding Author.}}
\email{lgbycl@cqu.edu.cn}
\affiliation{Department of Physics and Chongqing Key Laboratory for Strongly Coupled Physics, Chongqing University, Chongqing 401331, P. R. China}

\date{\today}

\begin{abstract}

We study stochastic gravitational wave production and baryon number generation at electroweak phase transition with the two Higgs doublet models. The produced stochastic gravitational wave during the strongly first-order phase transition can be probed by future space-based interferometers.
The {\it nonlocal} electroweak baryogengesis cannot address the observed Baryon asymmetry of the Universe successfully in the strongly first-order phase transition parameter spaces due to the CP violation phase is severely bounded by the electron electric dipole moment measurement ACMEII.

\end{abstract}

\maketitle

\section{Introduction}

 The Baryon asymmetry of the Universe is one of the fundamental and unsolved problems in particle physics and cosmology.
The electroweak baryogenesis (EWB), producing baryon asymmetry at the electroweak phase transition, is one of the most popular  mechanism to account for the Baryon asymmetry of the Universe due to potentially testability in future colliders and electric dipole moment measurements~\cite{Morrissey:2012db,Arkani-Hamed:2015vfh}. Recently, the gravitational wave study raises people's growing interest after the
observation of the binary black hole merger~\cite{Abbott:2016blz}, and the approval of the space-based detector LISA~\cite{Audley:2017drz}. The observation of a stochastic gravitational wave background produced at first-order phase transitions is one promising target of LISA~\cite{Klein:2015hvg}, since it can certainly provide important information on cosmology and behind high-energy physics, thus provide a novel opportunity to probe new physics beyond the standard model~\cite{Christensen:2018iqi,Mazumdar:2018dfl}.

The CP-violation beyond the standard model, as one crucial ingredient to address the BAU puzzle within EWB, can be tested by
the high precision electric dipole moment (EDM) measurements~\cite{Morrissey:2012db}.
The ACME further improved the sensitivity to the CP violation through the measure of electron EDM~\cite{Andreev:2018ayy}, which ruled out a lot new physics models addressing BAU with EWB~\cite{Cline:2017jvp}\footnote{The EDM measurements ruled out the {\it nonlocal} EWB with real or complex singlet models studied in Ref.~\cite{Jiang:2015cwa,Vaskonen:2016yiu}, with the dynamical CP violation scenario is an exception~\cite{Chao:2017oux,Bruggisser:2017lhc,Huang:2018aja,Ellis:2019flb,Wang:2019pet}.}.
 The conventionally adopt EWB mechanism with chiral transport equations is historically called {\it nonlocal } baryogenesis, which has been studied and developed extensively (See Ref.~\cite{Morrissey:2012db} for a recent review).
The {\it nonlocal} EWB through one-step electroweak phase transition (EWPT) usually requires a subsonic bubble wall velocity,
 i.e., typically $v_w\sim\mathcal{O}(10^{-2})$\footnote{We note that the wall velocity in the SM and MSSM has been estimated to be subsonic in Ref.~\cite{Moore:1995si,Moore:1995ua,John:2000zq}.}, to ensure there is enough time for the CP violating diffusion processes to generate chiral asymmetry ahead of the bubble which will be converted into net baryon asymmetry by the electroweak sphaleron in the symmetric phase \footnote{See Ref.~\cite{Cline:2020jre} for the case of two-step phase transition case, where EWB at high wall velocities has been studied.}.
 Meanwhile, previous studies of the bubble wall velocity taking into account the microphysics and hydrodynamics indicate that a large wall velocity corresponds to a strong phase transition~\cite{Moore:1995si,Moore:1995ua,Konstandin:2014zta,Dorsch:2018pat,Kozaczuk:2015owa,Huber:2011aa}, and a relativistic wall velocity is needed to produce a detectable gravitational wave signal from the EWPT~\cite{Caprini:2015zlo,Caprini:2019egz}.

In this work, we study the baryon number preservation criterion, the stochastic gravitational wave prediction at the phase transition in the Type-I and Type-II two Higgs doublet models, and the possibility to address the observed BAU with the {\it nonlocal} EWB during the EWPT process with these models.

\section{ Strongly first order phase transition}

The two Higgs doublet models (2HDMs) has been considered as a good and renormalizable framework to address BAU problem through {\it nonlocal} EWB~\cite{Fromme:2006cm,Huber:2006ri}. For the study of phase transition, we work within the CP-conserving 2HDMs since the CP violation allowed by the EDM experiments is believed to have negligible effects on the phase transition dynamics~\cite{Fromme:2006cm,Dorsch:2013wja,Andersen:2017ika}.
Theoretical and experimental constraints are imposed as in Ref.~\cite{Bernon:2017jgv}, where the type-II 2HDM suffers more server experimental constraints (especially the measurement of $B\to X_s \gamma$) in comparison with the type-I 2HDM scenario. Therefore, there are more parameter spaces in type-I 2HDM can achieve a first-order EWPT. To estimate the bubble nucleation situation, we adopt the first-order EWPT points with $v_c/T_c>1$ in both type-I and type-II 2HDM.

\begin{figure}[!htp]
\begin{center}
\includegraphics[width=0.22\textwidth]{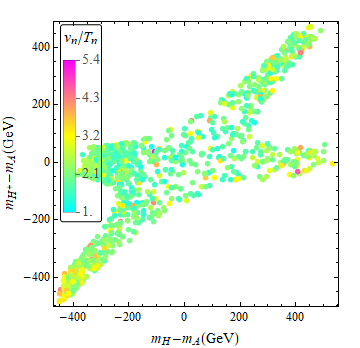}
\includegraphics[width=0.22\textwidth]{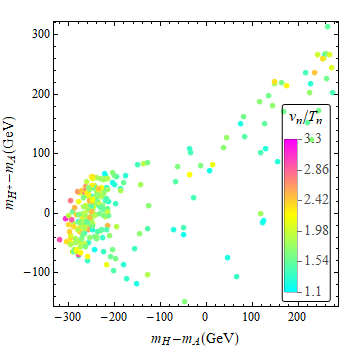}
\caption{The phase transition strength $v_n/T_n$ at nucleation temperature $T_n$ versus the mass difference $m_H-m_A$($m_{H^\pm} -m_A$). The left (right) panel shows the type-I (type-II) 2HDM scenario.}\label{pt}
\end{center}
\end{figure}

Our results are shown in Fig.~\ref{pt}.
The figure depicts that, there is more chance to achieve bubble nucleation with a relatively large phase transition strength in type-I 2HDM. The tendency of the figure reflects the electroweak precision bounds and the alignment assumptions~\cite{Bernon:2017jgv}, where a strong EWPT can be obtain in parameter spaces with large mass splitting, which is consistent with findings of Refs.~\cite{Dorsch:2013wja,Dorsch:2014qja,Basler:2016obg,Bernon:2017jgv}. As in previous studies, in this work we adopt the thermal field theory of 2HDM around electroweak scale and focus on EWPT. It should be note that in some parameter spaces of strong EWPT the large mass splitting corresponds to sizable quartic Higgs couplings which may suffer from Landau pole beyond the electroweak scale.

We now check the usually adopted strongly first-order phase transition condition coming from the requirement of baryon number preservation criterion (BNPC)~\cite{Patel:2011th,Morrissey:2012db}, which states that the electroweak sphaleron process inside the electroweak vacuum bubble (broken phase) should be sufficiently quenched~\cite{Dine:1991ck,Patel:2011th,Morrissey:2012db}. We define the quantity of $PT_{sph}$ as in Ref.~\cite{Zhou:2019uzq},
\begin{align}\label{eq:PTsph1}
PT_{sph}\equiv\frac{E_{\rm sph}(T)}{T} - 7 \ln \frac{v(T)}{T} + \ln \frac{T}{100 \rm{GeV}}\;.
\end{align}
The BNPC is met when the sphaleron rate in the broken phase is smaller than the Hubble expansion rate~\cite{Quiros:1999jp,Gan:2017mcv}\footnote{ Here, we note that, taking into account the bubble nucleation dynamics, one have a much stronger bound on the BNPC, which could be $\Gamma < 10^{-3} H$~\cite{Moore:2000jw}, and the duration of the phase transition can also alert the criteria given in Eq.~\ref{eq:PTsph2}~\cite{Moore:1998swa,Patel:2011th}. To settle down the prefactor of the $\Gamma_{sph}$, and therefore to lower the uncertainty of the criteria, lattice simulation of the Electroweak sphaleron inside the bubble is necessary. At present, the lattice simulation is not able to perform for the larger phase transition strength, and new method is required to cure the problem. }, which yields
\begin{align}\label{eq:PTsph2}
PT_{sph}> (35.9-42.8)\;.
\end{align}
The numerical range in the right hand side corresponds to the fluctuation determinant uncertainty~\cite{Dine:1991ck}, which is comparable with the uncertainty coming from the numerical lattice simulation of the sphaleron rate at the standard model electroweak {\it cross-over}~\cite{DOnofrio:2014rug}. We calculate the electroweak sphaleron energy $E_{\rm sph}(T)$ with the approach given in Appendix.~\ref{sec:app}.

 \begin{figure}[!htp]
\begin{center}
\includegraphics[width=0.215\textwidth]{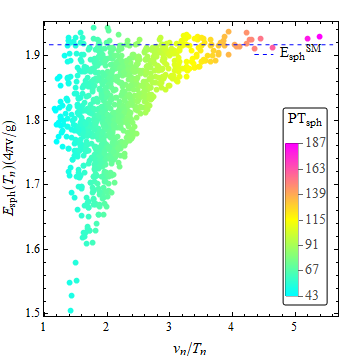}
\includegraphics[width=0.22\textwidth]{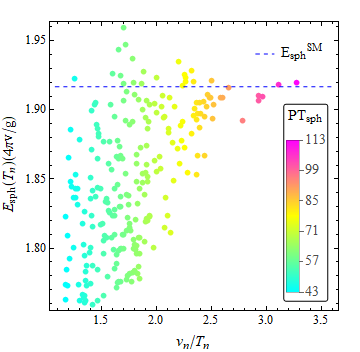}
\caption{The $PT_{sph}$ as a function of the phase transition strength $v_n/T_n$ and the electroweak sphaleron energy inside the broken-phase bubble. The left (right) panel shows the type-I (type-II) 2HDM scenario.}\label{bnpc}
\end{center}
\end{figure}

In Fig.~\ref{bnpc}, we present the relation among the electroweak sphaleron energy $E_{\rm sph}(T)$, the phase transition strength $v(T)/T$, and the quantity of $PT_{sph}$ inside the vacuum bubble. With the increase of the phase transition strength, the electroweak sphaleron energy at the nucleation temperature is found to close to the SM scenario ($E_{sph}^{SM}\sim 1.91 \times 4\pi v/g$), where one have a large $PT_{sph}$ which can sufficiently quench the electroweak sphaleron process inside the vacuum bubble and therefore keep the net baryon numbers generated at the EWPT. As will be studied in the following, one may have a strong gravitational wave signal that can be detected at LISA for a large $PT_{sph}$, meaning that the gravitational wave may serve as a test of parameter spaces for the EWB\footnote{This is consistent with the tree-level driven phase transition scenario~\cite{Zhou:2019uzq}. }.

\section{Gravitational Wave}
To predict the gravitational wave spectrum from the first-order EWPT, we first calculate the phase transition strength as $\alpha \equiv ({\bar \theta_s}(T_+) - {\bar \theta_b}(T_+))/3\omega_s $, with ${\bar \theta} \equiv e - p/c^2_{s}$~\cite{Giese:2020rtr,Giese:2020znk}, where \{e, p, $\omega$, and $c_{s}$\} is \{the energy density, the pressure, the enthalpy density, and the sound speed\} respectively. The subscripts $``b,-"$/$``s,+"$ indicate the quantities inside/outside the bubbles. With the help of the free energy density, the phase transition strength recast the form of~\cite{Guo:2021qcq}:
\begin{equation}
\alpha = \frac{1}{3 \omega_s}((1+\frac{1}{c^{2}_s}) \Delta V_{eff} - T\frac{d \Delta V_{eff}}{dT})|_{T=T_n}.
\end{equation}
where the $\Delta V_{eff}$ is the thermal effective potential difference between the symmetric and broken phase.
The other crucial parameter for the calculation of gravitational wave spectrum is $\beta$, which characterizes the inverse time duration of the phase transition and is defined as
\begin{equation}
\beta/H_n=T d (S_3(T)/T)/d T|_{T=T_n}\;,
\end{equation}
Where, $H_n$ is the Hubble constant at the nucleation temperature $T_n$. Both the two parameters ($\alpha$ and $\beta/H_n$) are evaluated after solving the bounce at the bubble nucleation temperature $T_n$~\cite{Masoumi:2017trx,Wainwright:2011kj,Athron:2019nbd}.

\begin{figure}[!htp]
\begin{center}
\includegraphics[width=0.22\textwidth]{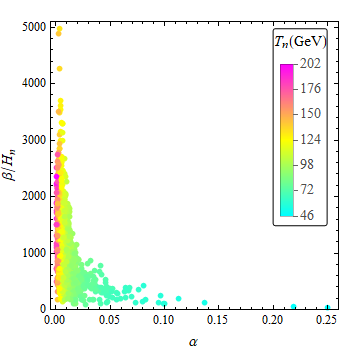}
\includegraphics[width=0.22\textwidth]{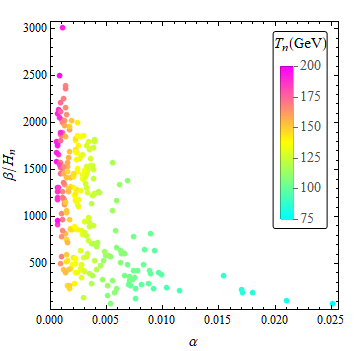}
\caption{ The GW parameters of $T_n$, $\beta/H_n$ and $\alpha$ in the type-I (left) and type-II (right) 2HDMs.}\label{GWpar}
\end{center}
\end{figure}

In Fig.\ref{GWpar}, we present the three crucial parameters for the gravitational wave predictions from electroweak phase transition with the 2HDMs, i.e., ($\alpha$,$\beta/H_n$,$T_n$).  The two plots depict that: 1)  With the decrease of the nucleation temperature $T_n$, one have decrease of $\beta/H_n$ and increase of $\alpha$, where one may have a large magnitude of gravitational wave signal from the electroweak phase transition;
2) In comparison with type-II 2HDM, type-I 2HDM allows a larger possibility to achieve a large $\alpha$ and low $\beta/H_n$ with low bubble nucleation temperature.

Another crucial parameter for the gravitational wave prediction from the phase transition is the wall velocity.
We checked the B\"odecker-Moore criterion~\cite{Bodeker:2009qy} and found that there is no runaway of the bubbles for these strongly first-order EWPT points.
Generally, the detectability of the gravitational waves from the phase transition at LISA requires a relativistic or ultra-relativistic bubble walls\cite{Caprini:2015zlo,Mazumdar:2018dfl}. By taking the wall velocity as a free parameter,
 we calculate the gravitational wave prediction from the electroweak phase transition in 2HDMs\footnote{See Ref.~\cite{
Hall:2019ank,Hall:2019rld} for the gravitational wave signals coming from a first order "electroweak" phase transition with 2HDM in the dark sector.}. Here, we consider the sound waves in the plasma~\cite{Hindmarsh:2013xza,Hindmarsh:2015qta} and
the magnetohydrodynamic turbulence (MHD)~\cite{Hindmarsh:2013xza,Hindmarsh:2015qta} that are believed to dominate the gravitational wave production during the phase transition, with the energy density spectrum from the sound waves simulated by the sound-shell model~\cite{Hindmarsh:2015qta}.

\begin{table}[!h]
\centering
\begin{tabular}{cccccccccccccccccccc}
\hline\hline
type I/II &  $m_H$  &  $m_{H^\pm}$  &   $m_A$  &  $\tan\beta$ & $\alpha$ & $\beta/H_n$ & $T_n$  \\
\hline
I/ BM 1  & 302.88 &   447.09  &    409.53     & 22.05 & 0.25 & 31.74 & 45.90\\ [+1mm]
\hline
 I/ BM 2  & 236.41 &  423.71   &    455.47     & 24.48 & 0.22 & 57.83 & 47.43\\ [+1mm]
 \hline
 II/BM 1 &    365.28  & 668.46  &   631.92   &  2.43& 0.03 & 69.65 & 75.29 \\ [+1mm]
\hline
II/BM 2  &  300.60  & 599.09   &   607.37     &   4.92& 0.02 & 104.62 & 79.26\\ [+1mm]
\hline
 II/BM 3 &  307.88  & 585.57   &   629.26     &   6.19& 0.02 & 189.16 & 82.63 \\ [+1mm]
\hline
II/BM 4  &   381.12  & 590.25  &   660.07     &  0.88& 0.02 & 193.69& 83.65\\ [+1mm]
\hline \hline
\end{tabular}
\def\baselinestretch{1.1}
\caption{Benchmarks in the type-I and type-II 2HDM in Fig.~\ref{gw} with Higgs masses are shown in units of GeV. }
\def\baselinestretch{1.0}
\label{tab:BP}
\end{table}

\begin{figure}[!htp]
\begin{center}
\includegraphics[width=0.4\textwidth]{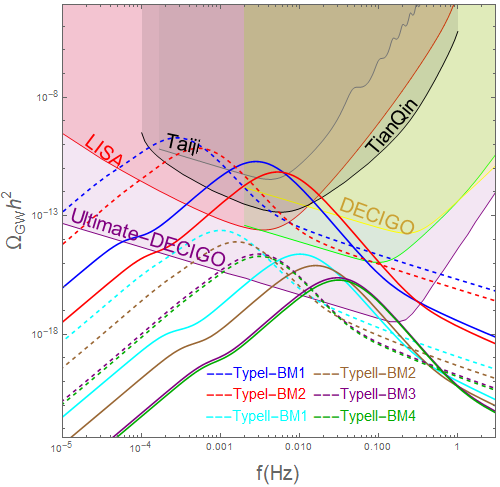}
\caption{Gravitational waves from the strong electroweak phase transition for benchmarks shown in Table.~\ref{tab:BP}. The solid (dashed) line depicts the $v_w=0.1 (1)$ scenarios.
}\label{gw}
\end{center}
\end{figure}

Our gravitational wave predictions for benchmarks given by the Table.~\ref{tab:BP} are shown in Fig.~\ref{gw}. Where the sound wave dominates all the gravitational wave sources and the wall velocity determines the amplitude and the peak of the sound wave spectrum. A lower wall velocity leads to a lower magnitude and a lower frequency of the peak of the gravitational wave spectrum, while it is necessary for the {\it nonlocal} EWB. The wall width would be thickness for the relevant {\it nonlocal} baryogenesis, and supercooling case with large $\alpha$ embracing thin wall does not happen in this study.

\section{Comment on electroweak baryogenesis}

We now study the realization of EWB at the EWPT. We focus on the {\it nonlocal} EWB situation under the current EDM experiments.
During a strongly first-order EWPT, the effective top quark mass embraces a space-time dependent phase varying across the slow bubble wall\footnote{As for the bottom quark scenario we refer to Ref~\cite{Liu:2011jh}.} and thus leads to a CPV source term~\cite{Huet:1995sh,Lee:2004we}
\begin{eqnarray}
S_t(z) \approx \frac{3}{2\pi^2} \left( \frac{m_t}{v \sin\beta} \right)^2 v_T^2(z) \theta'(z) v_w T \;.\label{eq:scpv}
\end{eqnarray}
Which relies on the bubble wall velocity and wall width calculation based on microphysics and hydrodynamics, and
the VEV's phase across the wall $\Delta\theta$ which depends on phase transition dynamics when the CP violation shows up. The CPV phase $\Delta\theta$ is supposed to have a similar size with the zero temperature phase $\xi$~\cite{Fromme:2006cm,Cline:2011mm}, i.e., $\Delta \theta \approx \alpha_b\sim 0.1$ is required to ensure enough BAU generation~\cite{Shu:2013uua}.
The current upper limit set by the current electron EDM measurement of ACME II is: $|d_e| < 1.1 \times 10^{-29} $ e cm at 90\% confidence level~\cite{Andreev:2018ayy}. To evade the server bounds, the cancellation mechanism~\cite{Shu:2013uua,Bian:2014zka,Bian:2016zba} is necessary.
On the other hand, a strong phase transition prefers a large mass splitting between the heavy CP-even Higgs and the heavy CP-odd Higgs as shown above, there wouldn't be cancellation driven by the mixing of the CP-odd and CP-even heavy Higgses due to the large mass splitting~\cite{Bian:2016zba}.
Therefore, the CP-violation bounds from the current electron EDM measurement may extremely reduce the possibility to achieve the correct BAU with the conventionally and extensively studied {\it nonlocal} EWB~\cite{Dorsch:2016nrg,Cline:2011mm,Cline:2017jvp}\footnote{Previous studies show that large scalar quartic couplings and narrow wall width is necessary for successful {\it nonlocal} EWB in the 2HDM, the lattice simulation is not valid anymore~\cite{Andersen:2017ika,Kainulainen:2019kyp} in the situation.}.
Considering gauge invariant Barr-Zee diagram contributions~\cite{Abe:2013qla}, we estimate the electron EDM by adopting the formula collected in Ref.~\cite{Bian:2014zka,Bian:2016awe,Inoue:2014nva,Bian:2017jpt} for the benchmarks in Table.~\ref{tab:BP} with the CP violation amplitude characterized by $\alpha_b$. Our estimation turns that $\alpha_b\leq 10^{-5}$, which is far small to yield the correct BAU with the {\it nonlocal} EWB.

\section{ Concluding remark}

Utilizing 2HDM, we checked the baryon number preservation criterion and found that a strong phase transition required by a strong stochastic gravitational wave corresponds to a sufficient quench of the electroweak sphaleron process as needed by the {\it nonlocal} EWB. We found that, in the type-I and type-II 2HDMs, the sufficient strong EWPT parameter spaces can be probed by the stochastic gravitational wave search conducted by the projected space mission LISA, TianQin~\cite{Luo:2015ght}, Taiji~\cite{Guo:2018npi}, Big Bang Observer (BBO)\cite{Yagi:2011wg}, and DECi-hertz Interferometer Gravitational wave Observatory (DECIGO)\cite{Kawamura:2011zz}.
For these parameter spaces, the large mass-splitting among heavy Higgses are required by the BNPC. The CP violation allowed by the eEDM experiment is too small to yield the correct BAU with {\it nonlocal} EWB.

The exact calculation of bubble wall velocity with microphysics and hydrodynamics and the lattice simulation of the electroweak sphaleron rate in the symmetric and broken phase during the phase transition are the two crucial ingredients to settle down if the observed BAU can be explained with the EWB.

\section{Acknowledgement}

Ligong Bian was supported by the National Natural Science Foundation of China under the grants Nos.12075041, 12047564, and the Fundamental Research Funds for the Central Universities of China (No. 2021CDJQY-011 and No. 2020CDJQY-Z003), and Chongqing Natural Science Foundation (Grants No.cstc2020jcyj-msxmX0814).
We are grateful to Guy David Moore, Michael J. Ramsey-Musolf, Mark Harmadsh,  Archil Kobakhidze, David E. Morrissey, Antonio Riotto, Michael Dine, Thomas Konstandin, Mark Trodden, Lauri Niemi, Kari Rummukainen, Jonathan Kozaczuk, Wouter Dekens, Xucheng Gan, Jiang-Hao Yu and Huai-Ke Guo for conversation and enlightenment discussions.

\appendix
\section{Electroweak sphaleron}
\label{sec:app}

The Electroweak sphaleron in the CP-violation 2HDM at zero temperature has also been studied previous at zero temperature, see Ref~\cite{Grant:2001at,Kastening:1991nw}. Since the CP-violation operators have ${\it null}$ contributions to the sphaleron energy~\cite{Gan:2017mcv} and also phase transition dynamics as previous studies, we therefore perform the Electroweak sphaleron calculations inside the vacuum bubble with CP-conserving 2HDM.
In this study, we employ the spherically symmetric ansatz and conduct the Electroweak sphaleron energy closely following the approach given in Ref.~\cite{Klinkhamer:1984di,Manton:1983nd}.
The configuration space are spanned by the following functions:
\begin{eqnarray}
 A_i(r,\theta,\phi)
 &=&-\frac{i}{g}f(r)\partial_i U(\theta, \phi) (U(\theta, \phi))^{-1},\\
\Phi_1(r,\theta,\phi)
&=&\frac{v_1}{\sqrt{2}}h_1(r)U(\theta, \phi)
	\left(
	\begin{array}{c}
	0 \\
	1
	\end{array}
	\right), \\
\Phi_2(r,\theta,\phi)
&=&\frac{v_2}{\sqrt{2}}h_2(r)U(\theta, \phi)
	\left(
	\begin{array}{c}
	0 \\
	1
	\end{array}
	\right),
\end{eqnarray}
where $A_i$ are SU(2) gauge fields, $A_{i}=\frac{1}{2} A_{i}^{a} \tau^{a}$, $v=\sqrt{v_1^2+v_2^2}$, and $U(\theta, \phi)$ is defined as
\begin{eqnarray}
U(\theta, \phi)=
	\left(
	\begin{array}{cc}
	\cos\theta & {e^{i \phi}  \sin \theta} \\
	{-e^{-i \phi}  \sin \theta} & \cos\theta
	\end{array}
	\right),
\end{eqnarray}
Adopting the $A_0 = 0$ gauge, the Electroweak sphaleron energy function can be obtained as:
\begin{eqnarray}
E_{\rm sph}[f,h_1,h_2]
&= \frac{4\pi v}{g}\int^\infty_0 d\xi~
\bigg[
	4\left(\frac{df}{d\xi}\right)^2+\frac{8}{\xi^2}(f-f^2)^2\nonumber\\
	&+\frac{\xi^2}{2}\frac{v_1^2}{v^2}\left(\frac{dh_1}{d\xi}\right)^2
	+\frac{\xi^2}{2}\frac{v_2^2}{v^2}\left(\frac{dh_2}{d\xi}\right)^2\nonumber\\
&+\bigg(\frac{v_1^2}{v^2}h_1^2+\frac{v_2^2}{v^2}h_2^2\bigg)(1-f)^2 \nonumber\\
&\hspace{0cm}
	+\frac{\xi^2}{g^2v^4}V(h_1, h_2)
\bigg],\label{Esph}
\end{eqnarray}
where $\xi=g v r$. From Eq.~(\ref{Esph}), the equations of motion are found to be
\begin{eqnarray}
&&\frac{d^2f}{d\xi^2}
= \frac{2}{\xi^2}f(1-f)(1-2f)-\bigg(\frac{v_1^2}{4v^2}h_1^2+\frac{v_2^2}{4v^2}h_2^2\bigg)\nonumber\\
&&\hspace{1cm}\times(1-f),\\
&&\frac{d}{d\xi}\left(\xi^2\frac{dh_1}{d\xi}\right)
= 2h_1(1-f)^2+\frac{\xi^2}{g^2v_1^2v^2}\frac{\partial V}{\partial h_1}, \\
&&\frac{d}{d\xi}\left(\xi^2\frac{dh_2}{d\xi}\right)
= 2h_2(1-f)^2+\frac{\xi^2}{g^2v_2^2v^2}\frac{\partial V}{\partial h_2}.
\end{eqnarray}
with following boundary conditions should be satisfied:
\begin{eqnarray}
&&\lim_{\xi\to0} f(\xi) = 0,\hspace{0.15cm} \lim_{\xi\to0} h_1(\xi) = 0,\hspace{0.15cm} \lim_{\xi\to0} h_2(\xi) = 0,  \\
&&\lim_{\xi\to\infty} f(\xi) = 1, \lim_{\xi\to\infty} h_1(\xi) = 1, \lim_{\xi\to\infty} h_2(\xi) = 1.
\end{eqnarray}

\begin{figure}[!htp]
\begin{center}
\includegraphics[width=0.4\textwidth]{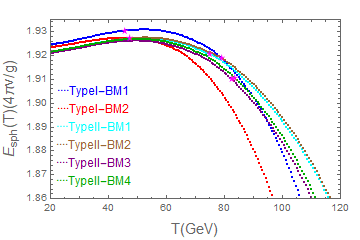}
\caption{The electroweak sphaleron energy
evolution with the temperature drops, these benchmark points are from Table.~I in the main text. }\label{sphT}
\end{center}
\end{figure}

The Electroweak sphaleron energy at finite temperature during the phase transition process ($E_{sph}(T)$) can be obtained with the replacement of the prefactor of $\frac{4\pi v}{g}$ of $E_{\rm sph}[f,h_1,h_2]$ given in Eq.\ref{Esph}: $v\rightarrow v(T_n)$~\cite{Patel:2011th}.
In Fig.~\ref{sphT}, we show the Electroweak sphaleron energy as a function of temperature, with temperature $T\leq T_c$.  It demonstrate that the
Electroweak sphaleron process would be highly suppressed in low temperatures after symmetry broken.

\bibliographystyle{utphys}
\bibliography{mybib}

\end{document}